\documentclass[preprint2]{aastex}

\def\url#1{{\ttfamily\def\/{/\discretionary{}{}{}}#1}}

\def\kms{\mbox{km/s}}

\def\kpc{\mbox{kpc}}
\def\kpch{\mbox{$h^{-1}$kpc}}

\def\Mpch{\mbox{$h^{-1}$Mpc}}
\def\Msun{\mbox{M$_\odot$}}

\def\Msunh{\mbox{$h^{-1}$M$_\odot$}}

\def\LCDM{{\char'3CDM}}
\def\mathnew{\mathsurround=0pt}
\def\simov#1#2{\lower .5pt\vbox{\baselineskip0pt
    \lineskip-.5pt\ialign{$\mathnew#1\hfil##\hfil$\crcr#2\crcr\sim\crcr}}}  
\def\simgreat{\mathrel{\mathpalette\simov >}}
\def\simless{\mathrel{\mathpalette\simov <}}
\def\'#1{\ifx#1i{\accent"13\i}\else{\accent"13#1}\fi}

\shorttitle{Halos in dynamical Dark Energy Models}
\shortauthors{Klypin et al.}
\begin{document}
\title{Halo properties in models with dynamical Dark Energy}
\author{A. Klypin}
\affil{Astronomy Department, New Mexico State University, Box 30001, Department
4500, Las Cruces, NM 88003-0001}

\author{A.V. Macci\`o, R. Mainini \& S.A. Bonometto}
\affil{Physics Department G. Occhialini, Universit\`a degli Studi di
Milano--Bicocca, Piazza della Scienza 3, I20126 Milano (Italy)}
\affil{I.N.F.N., Via Celoria 16, I20133 Milano (Italy)}

\begin{abstract}
We study properties of dark matter halos in a variety of models which
include Dark Energy (DE). We consider both DE due to a scalar field
self--interacting through Ratra--Peebles or SUGRA potentials, and DE
with constant negative $w=p/\rho > -1$. We find that at redshift zero
the nonlinear power spectrum of the dark matter, and the mass function
of halos, practically do not depend on DE state equation and are
almost indistinguishable from predictions of the \LCDM~ model. This is
consistent with the nonlinear analysis presented in the accompanying
paper. It is also a welcome feature because
\LCDM~models fit a large variety of data. On the other hand, at high
redshifts DE models show substantial differences from \LCDM~and
substantial differences among themselves. Halo profiles differ even
at $z=0$. DE halos are denser than \LCDM~ in their central parts
because the DE halos collapse earlier.  Nevertheless, differences
between the models are not so large. For example, the density at
10~\kpc~ of a DE $\sim 10^{13}\Msun$ halo deviates from \LCDM~ by not
more than 50$\, \%$. This, however, means that DE is not a way to ease
the problem with cuspy dark matter profiles. Addressing another
cosmological problem - abundance of subhalos -- we find that the
number of satellites of halos in various DE models does not change
relative to the \LCDM, when normalized to the same circular velocity
of the parent halo.  To summarize, the best way to find which DE model
fits the observed Universe is to look for evolution of halo
properties.  For example, the abundance of galaxy groups with mass
larger than $10^{13}\Msunh$ at $z\simgreat 2$ can be used to
discriminate between the models, and, thus, to constrain the nature of
DE.

\end{abstract}


\section{Introduction}

Mounting observational evidence for Dark Energy
\citep{Perlmutter, Riess, Tegmark01, Netterfield, Pogosian03,
Efstathiou2, Percival, spergel2003},
which probably contributes $\sim 70\%$ of the critical density of the
Universe, rises a number of questions regarding consequences for
galaxy formation. Traditionally, DE is described by the parameter
$w=p/\rho$ characterizing its equation of state. The \LCDM~~
model ($w=-1$) was extensively studied during the last decade. 
Models with a constant negative $w > -1$ were much less studied, let
alone physically motivated models with variable $w$
\citep{Mainini03a, Mainini03b}, 
for which no N--body simulation has been performed yet. 
Observations \citep{spergel2003,Sch2003a} 
limit the present day value of $w \simless 
-0.8$, though the limit has been derived for constant--$w$ models only.

In the accompanying paper \citep{Mainini03b} we describe procedures
and give approximations for different quantities encountered in the
linear and nonlinear analyzes of fluctuations in models in which DE is
produced by a self--interacting scalar field (dynamical DE). In this
paper we use these approximations to perform N--body simulations of
models with dynamical DE and to study different properties of dark
matter halos in such $N-$body simulations. For completeness we also
study models with constant $w =-0.6$ and $w =-0.8$

Our main interest is in the models with varying $w$. These models use
physically motivated potentials of scalar field and admit tracker
solutions.  We focus on the two most popular variants of dynamical DE
\citep{wett1, RP, wett2}.  The first model was proposed by \citet[][RP
hereafter]{RP}. It produces rather slow evolution of $w$. The second
model \citep{BraxMartin99, BraxMartinRiazuelo, BraxMartin00} is based
on simple potentials in supergravity (SUGRA). It results in much faster
evolving $w$.
Hence, RP and SUGRA potentials cover a large spectrum of evolving
$w$. These potentials are written as
\begin{eqnarray}
V(\phi) &=& \frac{\Lambda^{4+\alpha}} {\phi^\alpha} \qquad RP, \\
V(\phi) &=& \frac{\Lambda^{4+\alpha}}{\phi^\alpha} \exp (4\pi G \phi^2)~~~ SUGRA.
\end{eqnarray}
Here $\Lambda$ is an energy scale, currently set in the range
$10^2$--$10^{10}\, $GeV, relevant for fundamental interaction
physics. The potentials depend also on the exponent $\alpha$.  The
parameters $\Lambda$ and $\alpha$ define the DE density parameter
$\Omega_{DE}$. However, we prefer to use $\Lambda$ and $\Omega_{DE}$
as independent parameters.  Figure~10 in \citet{Mainini03b} gives
examples of $w$ evolution for RP and SUGRA models. The RP model
considered in this paper has $\Lambda=10^3$ GeV. At redshift $z=0$ it has
$w=-0.5$. The value of $w$ gradually changes with the redshift: at
$z=5$ it is close to $-0.4$. The SUGRA model has $w=-0.85$ at $z=0$,
but $w$ drastically changes with redshift: $w\approx -0.4$ at $z=5$.
Although the $w$ interval spanned by the RP model covers values
significantly 
above 
-0.8 (not favored by observations), this case is
still important both as a limiting reference case and to emphasize
that models with constant $w$ and models with variable $w$ produce
different results even if average values of $w$ are not much
different.  Constant $w$ models have no physical motivation and can
only be justified as toy models to explore the parameter space. The
typical values of $w$ observed in dynamical DE models, however,
suggest to use $w= -0.8$ and $w= -0.6$ for the models with constant
$w$.

\section{Simulations}
The Adaptive Refinement  Tree code (ART;  Kravtsov, Klypin \& Khokhlov
1997) was used   to run the  simulations.  The ART code starts  with a
uniform  grid, which  covers the whole   computational box. This  grid
defines  the  lowest (zeroth) level of   resolution of the simulation.
The standard Particles-Mesh algorithms are used to compute density and
gravitational  potential  on the   zeroth-level  mesh.  The  ART  code
reaches   high force resolution by refining   all high density regions
using   an  automated  refinement   algorithm.   The   refinements are
recursive:  the refined regions can  also be  refined, each subsequent
refinement having half of the previous level's cell size. This creates
a hierarchy of refinement  meshes  of different resolution, size,  and
geometry  covering regions of  interest. Because each individual cubic
cell can be refined, the shape of the refinement mesh can be arbitrary
and match effectively the geometry of the region of interest.

The criterion for refinement is the local density of particles: if the
number of particles in a mesh cell (as  estimated by the Cloud-In-Cell
method)  exceeds  the   level  $n_{\rm thresh}$,  the   cell  is split
(``refined'')  into  8   cells of   the   next refinement  level.  The
refinement threshold may depend on the refinement level. The code uses
the   expansion  parameter  $a$ as  the    time variable.   During the
integration,   spatial  refinement is    accompanied  by  temporal
refinement.  Namely, each level of refinement, $l$, is integrated with
its own time  step $\Delta a_l=\Delta  a_0/2^l$, where $\Delta a_0$ is
the global time step  of the zeroth  refinement level.   This variable
time stepping  is very important for  accuracy of the results.  As the
force resolution  increases, more  steps  are needed to integrate  the
trajectories accurately.  Extensive  tests of the code and comparisons
with other  numerical $N$-body codes can  be found  in Kravtsov (1999)
and Knebe et al. (2000).
The code was modified to handle DE of different types.

A large number of simulations were performed. The simulations have
different  sizes of computational box, different force and mass
resolutions. Table~1 lists parameters of all our simulations.  This
large set of simulations allows us to study properties of halos
ranging from dwarf satellites to clusters of galaxies.
All simulations were extensively studied. We find that in all cases
the results are bracketed by the \LCDM~and the RP models. Normally,
differences between models are not very large. In order to avoid
too crowded plots, in most of the presented plots we show only results of
these two models.

\begin{deluxetable}{lllllll}
\tablecolumns{6} \tablewidth{0pc} \tablecaption {Parameters of
simulations} \tablehead{ \colhead{Model} & \colhead{$\sigma_8$} &
\colhead{Box size} & \colhead{Number of particles} & \colhead{Mass
resolution}& \colhead{Force resolution} \\ & &(\Mpch) & & (\Msunh) &
(\kpch) \\} \startdata $w=-0.6$ & 0.75 & 80 & 128$^3$ & $2.0\times
10^{10}$ & 5 \\ $w=-0.8$ & 0.75 & 80 & 128$^3$ & $2.0\times 10^{10}$ &
5 \\ RP & 0.75 & 60 & 128$^3$ & $8.4\times 10^{9}$ & 5 \\ &0.75 & 80 &
128$^3$ & $2.0\times 10^{10}$ & 5 \\ &0.75 & 160 & 256$^3$ &
$2.0\times 10^{10}$ & 10 \\ &0.75 & 80 & $7.32\times 10^5$ &
$3.1\times 10^{8}$ & 1.2 \\ &1.00 & 60 & $7.32\times 10^5$ &
$1.3\times 10^{8}$ & 0.9 \\ SUGRA & 0.75 & 60 & 128$^3$ & $8.4\times
10^{9}$ & 5 \\ &0.75 & 80 & 128$^3$ & $2.0\times 10^{10}$ & 5 \\ &0.75
& 160 & 256$^3$ & $2.0\times 10^{10}$ & 5 \\ \LCDM~& 0.75 & 60 &
128$^3$ & $8.4\times 10^{9}$ & 5 \\ &0.75 & 80 & 128$^3$ & $2.0\times
10^{10}$ & 5 \\ &0.75 & 160 & 256$^3$ & $2.0\times 10^{10}$ & 20 \\
&0.75 & 80 & $7.32\times 10^5$ & $3.1\times 10^{8}$ & 1.2 \\ &1.00 &
60 & $7.32\times 10^5$ & $1.3\times 10^{8}$ & 0.9 \\

\enddata
\label{tab4}
\end{deluxetable}


\section{Statistics of halos: power spectrum,  mass and velocity functions}

\begin{figure}[tb!]
\plotone{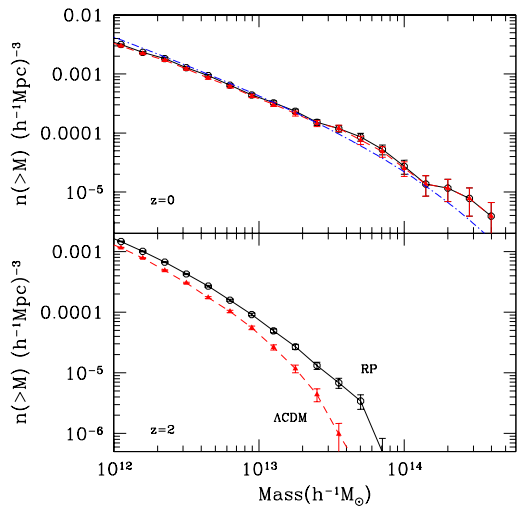}
\caption{\small The mass function of isolated halos in the \LCDM~ and
RP models. The masses are found within virial radii. Results of all
the rest of the models are in between these two models. At $z=0$ there
is no difference between the mass functions. The dot-dashed curve
shows ST prediction. At $z=2$ the mass functions of $w\neq -1$
models are above that of the \LCDM.}
\label{fig:massz0}
\end{figure}

Figure~\ref{fig:massz0} shows the mass function for isolated halos in
the RP and the \LCDM~ models. The simulations have the same initial
phases and the same value $\sigma_8=0.75$.  Thus, the differences
between models are only due to different $w(t)$.  Remarkably, at $z=0$
the mass functions are practically indistinguishable: a mass function
has no ``memory'' of the past evolution. In this figure we show only
two models, but all other models show the same results at $z=0$. The
mass function is well fitted by the approximation provided by Sheth \&
Tormen \citep[ST,][]{ShethTormen99, Sheth01, ShethTormen02}.

At higher redshifts the situation is quite different: mass functions
deviate substantially. Bottom panel in Figure~\ref{fig:massz0}
clearly demonstrates this: the number of clusters with mass large than
$\approx 3 \times 10^{13}\Msunh$ is almost ten times larger in the RP
simulation.  The differences depend on mass. They are larger for
massive clusters and much smaller for less massive halos. For
galaxy-size halos with mass $\sim 10^{12}\Msunh$ the differences are
only $\sim 20$\%, which will be difficult to detect observationally.

\begin{figure}[tb!]
\plotone{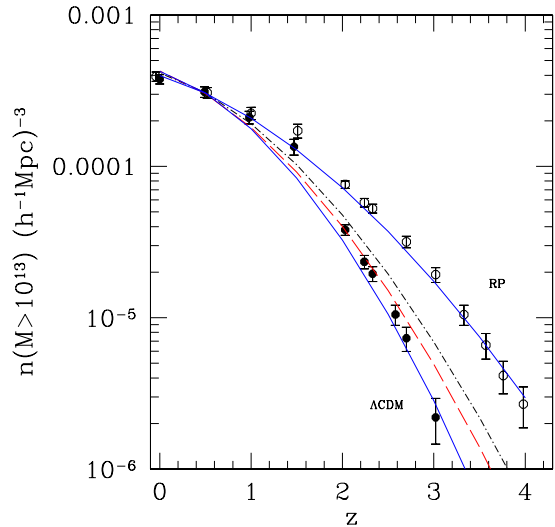}
\caption{\small The evolution of the number density of halos with
virial mass larger than $10^{13}\Mpch$ for different models.  The
bottom and top symbols are for $N-$body results in \LCDM~ and in RP
models. The curves show ST approximations. The dashed (dot-dashed)
curve is for $w=-0.6$ (SUGRA) models. There is hardly any difference between
models at redshifts smaller than $z=1$. At higher $z$ the number of
halos in \LCDM~ declines faster than for other models.}
\label{fig:massz}
\end{figure}

The dependence of halo abundance with redshift is further illustrated
in Figure~\ref{fig:massz}, where we study halos with mass of a group
of galaxies. There is almost no way to distinguish models at recent
times $z< 1$. But at $z=2-3$ the differences are quite significant. We
note that observational detection of group-size halos at high
redshifts is difficult, but feasible. We know how these objects should
look like - almost the same as nearby groups. A group at high redshift
should be more compact than a group at $z=0$ and it should consist of
3-10 Milky-Way size galaxies. Galaxies are expected to be distorted by
interactions. A sample of few thousands galaxies can be used to count
the number of groups. 
 Comparison with the number of groups at present
moment seems to be the way to discriminate between different
models of DE.

\begin{figure}[tb!]
\plotone{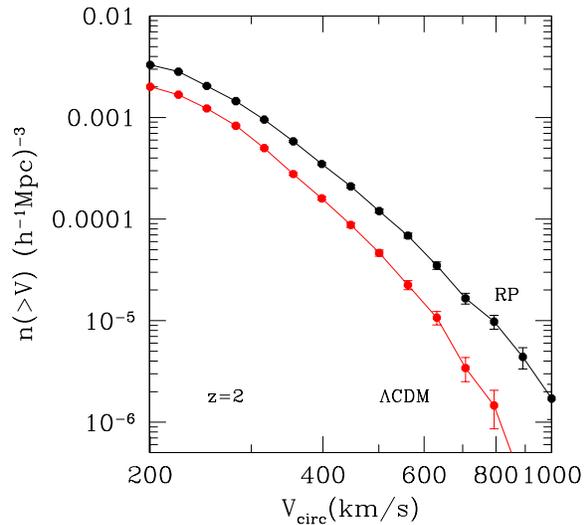}
\caption{\small The number density of halos with the maximum circular
velocity larger than $V_{\rm circ}$ at $z=2$. We show  RP and
\LCDM~ models.}
\label{fig:vc}
\end{figure}

For each halo we find the density profile and estimate the maximum
circular velocity $V_{\rm circ}=\sqrt{GM(<r)/r}$. We then construct
the circular velocity function of halos -- the number density of halos
with given $V_{\rm circ}$. The velocity function is
a kin of the mass
function, but it probes deeper inside halos. For a typical halo
discussed here with a concentration $C\approx 10$, the radius of the
maximum circular velocity is about five times smaller than the virial
radius. Figure~\ref{fig:vc} shows the velocity function at $z=2$ for
RP and \LCDM~ models. Just as in the case of the mass function, the
differences between models are larger at high redshifts. At given
redshift the differences are larger for massive halos. Still, the
velocity function brings new results. Even at $z=2$ the mass functions
are very close for low mass halos with virial mass $\approx
10^{12}\Msunh$. These halos have $V_{\rm circ}\approx 200~\kms$. The
velocity functions at that $V_{\rm circ}$ are visibly different: RP
model has about 1.5 times more halos. The only way to explain this is
to have more concentrated halos in RP model. In the next section we will
explore this possibility in detail.

\begin{figure}[tb!]
\plotone{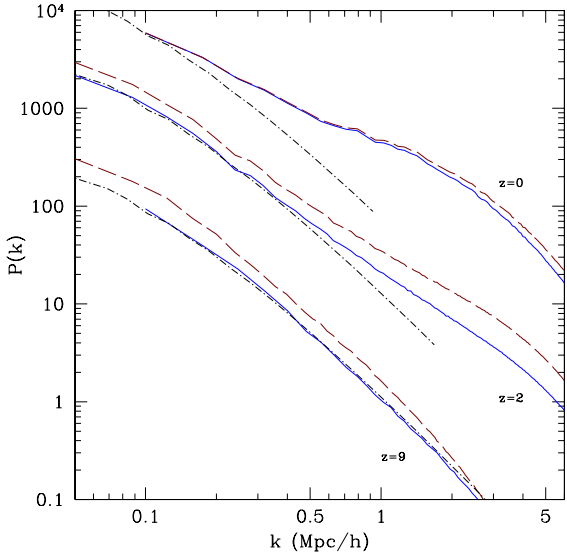}
\caption{\small Power spectrum of fluctuations of dark matter for
the \LCDM~ (full curves) and for the RP (dashed curves) models at different
redshifts. The dot-dashed curves show linear spectra for the \LCDM~
model.  At high redshifts fluctuations in the RP model are larger than
for the \LCDM~ model resulting in earlier collapse and in more dense
halos. The differences in $P(k)$ are the largest at $z=2-3$.
}
\label{fig:power}
\end{figure}

Figure~\ref{fig:power} shows the evolution of the power spectrum
$P(k)$ for fluctuations in the dark matter. The power spectrum
basically follows the same pattern as the mass function: relatively
large differences at high redshift, which become much smaller at
$z=0$. At $z=0$ the deviations remain only on small scales ($k> 2$).

\section{Halo structure }
\begin{figure}[tb!]
\plotone{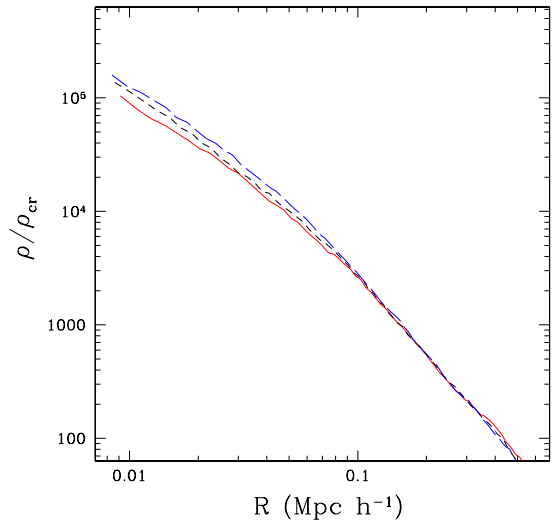}
\caption{\small Profile of the same halo simulated in different
models. The full curve is for the \LCDM~model. The shot (long) dashed
curve is for SUGRA (RP) models. The halo has the virial mass $5\times
10^{13}\Msunh$.  For different DE models the density profiles are
practically the same in the outer $R> 100\kpch$ region.  Each profile
is well approximated by the NFW profile.}
\label{fig:profile}
\end{figure}

We start our study of halo profiles by making high resolution
simulations of the same halo in different models. The halo was
initially identified in a low resolution run. Short waves were added
to the spectrum of initial perturbations and the halo was simulated
again using $\approx 2\times 10^5$ particles.  In the \LCDM~ model the
halo has virial mass $5\times 10^{13}\Msunh$ and virial radius
$730\kpch$. It is accurately fitted by the NFW profile 
\citep{Navarro97}
with the
concentration $C_{\rm vir}=7.2$. In the RP model the virial radius is
$680\kpch$ -- visibly smaller than for the \LCDM~ halo. The RP halo
also have larger maximum circular velocity as compared with the 
\LCDM~ halo.  Figure~\ref{fig:profile} shows profiles of the halo in
the \LCDM, RP, and SUGRA models.  In spite of the fact that the virial
radii for all the models are different, the density profiles in the
outer part of the halo $R>100\kpch$ are practically the same: from
$100\kpch$ to $700\kpch$ the differences are less than 10\%.  The
halos  differ only in the central region $R< 100\kpch$.  The RP
halo is clearly denser and more concentrated than the \LCDM~ halo with
the SUGRA halo being in between.  This difference can be used to
discriminate between the models. Yet, it will not be easy because the
differences are relatively small: factor 1.5 at $10\kpch$.

The RP has a smaller virial radius because the virial radius in the RP
model is defined at larger overdensity 
($\Delta_{vir,RP}$ = 149.8
$\rho_{cr}$).This is the
prediction of the top-hat model of halo collapse used to define the virial
mass \citep{Mainini03b}. 
There is nothing wrong with it, but it complicates the comparison of
density profiles and concentrations in different DE models.  For
example, a halo with exactly the same profile will have different
virial radii and, thus, different concentrations in different DE
models. In order to make comparison of density profiles less
ambiguous, we decided to measure the halo concentration as the ratio
of the radius at the overdensity of the \LCDM~ model (103 times the
critical density) to the characteristic (``core'') radius of the NFW
profile. The effect of using the radius at the constant overdensity
instead of the virial radius is relatively small. For typical RP halo
with virial mass $\sim 10^{13}\Msunh$ the virial radius is $\sim 15$\%
smaller as compared with the constant overdensity radius.

\begin{figure}[tb!]
\plotone{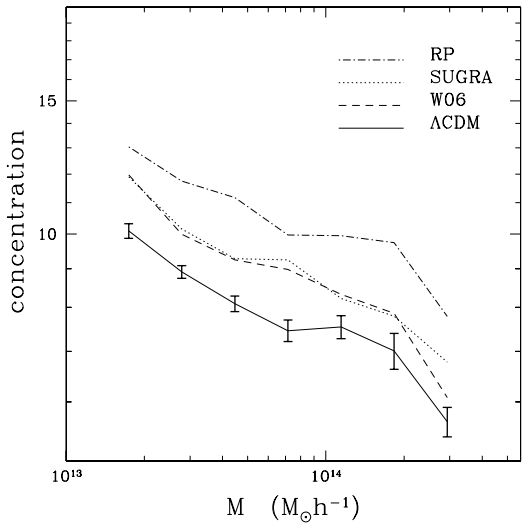}
\caption{\small Dependence of concentration on halo mass.
  Halos for models with $w\neq -1$ are all more concentrated and,
  thus, are denser than the halos in the \LCDM~ model. To avoid
  crowding we show statistical errors only for the \LCDM~ model.}
\label{fig:conc}
\end{figure}

\begin{figure}[tb!]
\plotone{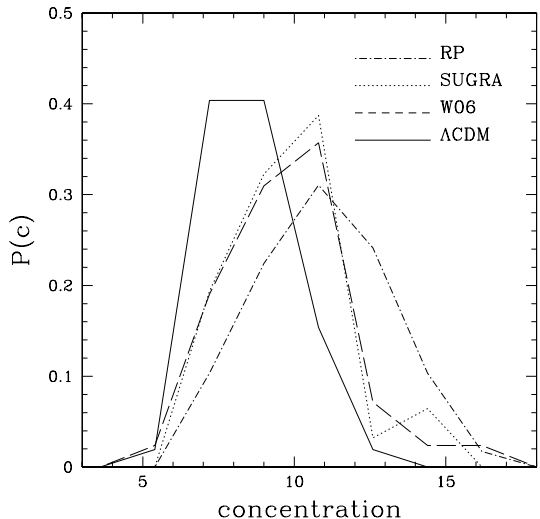}
\caption{\small Distribution of halo concentrations for halos in mass
range $(5-10)\times 10^{13}\Msunh$ for different models. Halos with
large deviations from NFW fits (non-relaxed halos) are not used.}
\label{fig:cprob}
\end{figure}

We also study profiles of hundreds of halos in simulations with lower
resolution.  Figure~\ref{fig:conc} shows the dependence of halo
concentration on the mass of halos in simulations with $80\Mpch$ box
with $\sigma_8=0.75$. This plot shows the same tendency, which we
found for the high-resolution halo: models with dynamical DE produce
more concentrated halos.  Figure~\ref{fig:cprob} shows the
distribution of halo concentrations for halos in mass range
$(5-10)\times 10^{13}\Msunh$. Halos with large
deviations from NFW fits (non-relaxed halos) are not used. The spread
of concentrations in the \LCDM~ model is about twice smaller than in
\citet{Bullock}.

\begin{figure}[tb!]
\plotone{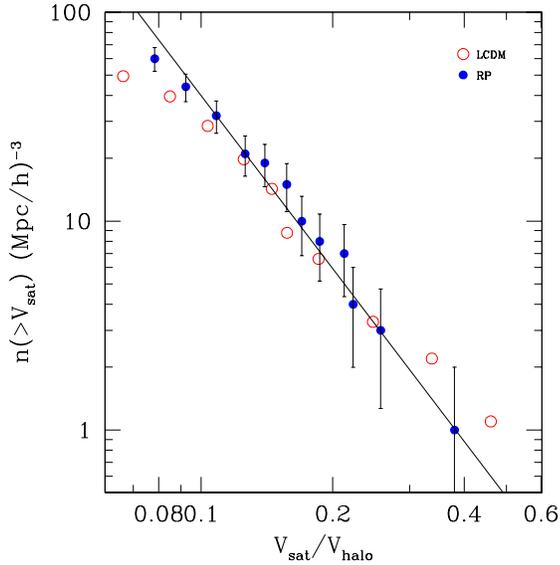}
\caption{\small Abundance of subhalos in a halo with virial mass
$M_{\rm vir}=2.4\times 10^{13}\Msunh$.  When normalized to the
circular velocity of the parent halo, the velocity function is the
same for both the RP and the \LCDM~models and is well approximated by
the power-law $n(>V)\propto V^{-2.75}$.  Vertical bars indicate the 
shot-noise errors.
}
\label{fig:satellites}
\end{figure}

Abundance of subhalos in the \LCDM~ model is a known problem
\citep{Klypin99, Moore99}.  It is interesting to find where dynamical
DE models stand regarding the problem. Because fluctuations in
dynamical DE models collapse earlier than in the \LCDM~ model, one
naively expects that the number of subhalos is also larger.  We study
the number of subhalos in a high resolution halo. The halo is
simulated in the RP and the \LCDM~ models. The halo with virial mass
$2.4\times 10^{13}\Msunh$ is resolved with particles of mass
$1.3\times 10^8\Msunh$. The maximum circular velocity of the halo in
\LCDM~ (RP) model is $522~\kms$ ($594~\kms$).  The force resolution
$\approx 1\kpch$ allows us to resolve dwarf DM halos with circular
velocity larger than $30~\kms$. For each (sub)halo we measure the
density profile and estimate the value of the maximum circular
velocity.

  The number of subhalos in the RP halo is larger than in the \LCDM~
halo: Inside the radius with the mean overdensity 103 of the critical
density there are 87 satellites in the RP halo and 52 satellites in
the \LCDM~ model.  Thus, there are a factor of 1.7 more satellites in
the RP halo.  Nevertheless, this large difference can be misleading
because the circular velocity of the RP halo is larger by factor 1.14
and halos with larger circular velocity have a tendency to have more
satellites \citep{Klypin99}. In Figure~\ref{fig:satellites} we plot
the number of satellites as the function of the ratio of the satellite
velocity to the halo velocity.  Differences between the models are
very small. It is also interesting to note that the velocity function
of the subhalos is well approximated by the power-law $n(>V)\propto
V^{-2.75}$. The slope of the power is the same as for subhalos of
Milky Way-size halos \citep{Klypin99}. In other words, it indicates
that the slope does not depend on the mass of halo and does not depend
on the DE equation of state.

\section{Discussion and conclusions }

Models with the dynamical DE are in infant state. We do not know the
nature of DE. Thus, 
a great arbitrariness exists on the choice 
of the equation of state $w(t)$.

At first sight it seems that the situation is
hopeless. 
This paper shows that this is not true: 
if we accept that $w$ is close to -1 at $z=0$, as many observations
suggest, and that $w$ monotonically increases with redshift, dynamical
models are useful and can produce definite predictions for properties
of halos and for galaxies hosted by the halos.  Furthermore, the
differences between rather extreme models of DE appear to be
relatively small. In other words, one can make detailed predictions
for properties of dark matter halos and for their clustering without
knowing too many details of $w$ evolution.  Yet, the differences
between models of DE exist and can be used to constrain the value and
the evolution of $w$.  In particular, distinguishing DE models
by using only the value
of $w$ at the present time is clearly insufficient.

The main tendency, which we find in all DE models is that halos tend
to collapse earlier. As the result, they are more concentrated and
more dense in the inner parts. Nevertheless, differences are not so
large. For example, the density at 10~\kpc~ of a  $\sim
10^{13}\Msun$ halo in a dynamical DE model deviates 
from \LCDM~ not more than by 50$\,\%$. 
This, however, means that DE is not a way to ease the problem
with cuspy dark matter profiles. 
Nevertheless, the differences in halo profiles can be exploited.
Denser cluster profiles in dynamical DE models can be tested by both
the weak (Bartelmann et al 2002) and especially by  the strong gravitational
lensing. Bartelmann et al (1998) and Meneghetti et al (2000) argue
that the arclet statistics favors \LCDM~models when compared with the
open CDM models. In this respect dynamical DE models are between the
above two models. This problem deserves further investigation.

We find that the best way to find which DE model fits the observed
Universe best is to look for evolution of halo properties. For example,
comparison of low- and high-z ($z\simgreat 2$) abundances of galaxy
groups with mass larger than $10^{13}\Msunh$ can be used to
discriminate between models. Potentially, clustering of galaxies at
redshifts $2-3$ can also be used for this.

In this paper we mostly pay attention to the group-size halos with
mass $\sim 10^{13}\Msunh$ at high redshifts as a probe for the DE.  In
the accompanying paper \citet{Mainini03b} we also argue that abundance
of clusters at intermediate redshifts can be used as a test for DE models.
Available cluster samples, unfortunately, still include too few clusters 
at intermediate and high redshift. 

To directly investigate the cluster mass
function at intermediate redshift with optical instruments, deep
optical or near  infrared data are used. Exploiting this kind of data the
Red--Sequence Cluster Survey \citep{Gladders2000} and the Las Campanas
Distant Cluster Survey \citep{Nelson2002} were compiled.
Taking carefully in to account selections effects is rather hard and
these samples include just tenth of objects.
Selection effects are
easier to handle for clusters detected in X--rays. The ROSAT data were
used to compile a number of cluster catalogs \citep{Ebeling1996,
Ebeling2000, degrandi}. The most numerous sample of flux 
limited clusters (REFLEX: Guzzo et al 1999,
Schuecker et al 2003b) is based on the ROSAT observations. It includes
426 objects with redshifts up to $z \sim 0.3$. The XMM Survey (Pierre
2000) will add another 800 clusters with redshifts up to $z \sim 1$.
Hopefully, follow--up optical programs will provide redshifts for the
clusters in the catalogs.
While designed for different
goals,  REFLEX  have been already used
to constrain many cosmological parameters such as $\sigma_8$, and,
together with SNIa data, it provides important constraints
on the DE equation of state (Schuecker et al 2003a).

The Suniaev-Zeldovich (SZ) effect (scattering of CMB photons by the
hot intracluster gas) is even more promising for detection of
high--$z$ clusters (La Roque et al, 2003; Weller et al 2002, Hu
2003). The shallow all-sky survey that the PLANCK experiment will
produce will be supplemented by narrower surveys covering a smaller 
fraction of the sky, based on
interferometric devices (OCRA: Browne et al 2000; SZA: Carlstrom et al
2000; AMIBA: Lo et al 2000; AMI: Kneisel 2001).

These new cluster catalogs require more extensive and detailed
theoretical modeling.  Confrontation of new observational data with
theoretical predictions will be able to discriminate between different
DE models.

In our analysis we also address another important issue: the abundance
of subhalos. It is well known \citep{Klypin99, Moore99} that in the
\LCDM~ model the number of predicted dwarf dark matter satellites
significantly exceeds the observed number of satellite galaxies in the
Local Group. There are different possibilities to explain this excess.
The most attractive explanation is related with the reionization of
the Universe resulting in heating of
gas in dwarf halos, which prevents them from becoming galaxies
\citep{Bullock2000, Bullock2001,Somerville2002,Benson2002}.

We find that the number of
satellites of halos, at $z=0$, in various DE models does not change relative to
the \LCDM, when normalized to the same circular velocity of parent halo.  
If the reionization of the Universe is the solution of the problem,
then the DE models predict an earlier reionization of the Universe,
because the earlier collapse of dwarf dark matter halos requires an
earlier reionization to avoid too many satellites at redshift zero.
The recent WMAP results (Kogut et al 2003, Spergel et al 2003), can be
interpreted as giving a large opacity for CMB photons $\tau \simeq
0.17 \pm 0.04$.  If true, this requires that the reionization occurred
at a redshift $z_{ri} \sim 13$--20, which is too large for the
standard \LCDM~ model \citep{Gnedin2000}. If the early reionization
happens in the \LCDM~ model, it would predict too few satellites
for the Local Group because too few dwarf halos collapse that early.
Models with SUGRA DE seem to be in a better position to fit WMAP
results and, at the same time, the observed number of satellites:
in fact, in this model, halos collapse at higher redshifts as compared
with the \LCDM~ model.

\section*{ACKNOWLEDGEMENTS}

We thank INAF for allowing us the CPU time 
to perform some of the simulations used in this work at the 
CINECA consortium (grant cnami44a on the SGI Origin 3800 machine).

\end{document}